\documentstyle[rotate,psfig]{laa}

\newcommand\lea{\mathrel{\raise .4ex\hbox{\rlap{$<$}\lower 1.2ex\hbox{$\sim$}}}}
\newcommand\gea{\mathrel{\raise .4ex\hbox{\rlap{$>$}\lower 1.2ex\hbox{$\sim$}}}}

\newcommand\cl{\centerline}
\renewcommand\deg{\ifmmode^\circ\else$^\circ$\fi}
\newcommand\solar{\ifmmode_{\mathord\odot}\else$_{\mathord\odot}$\fi}

\begin{document}

\thesaurus{13.18.1, 11.10.1, 11.17.3, 11.17.4 Cygnus A, 11.17.4 3C\,309.1,
11.17.4 3C\,345}

\title{Ultracompact jets in AGN}

\author{
A.P. Lobanov 
}

\offprints{A.P. Lobanov}

\institute{
Max-Planck-Institut f\"ur Radioastronomie, Auf dem H\"ugel 69, D-53121 Bonn, Germany }

\date{Received August 15; accepted September 24, 1997}

\maketitle

\begin{abstract}
We study the properties of ultracompact jets in several prominent
radio sources (Cygnus A, 3C\,309.1, 3C\,345, 3C\,395, 4C\,39.25, 
and 1038+528\,A), using the 
frequency dependence of observed position of the optically thick jet core.
Frequency dependent offsets of the core positions are used for
calculating the luminosities, magnetic fields, and geometrical properties of
the jets. Opacity effects are studied, for both synchrotron self--absorption
and external free--free absorption. Pressure and
density gradients in the jets and in surrounding ambient medium are shown to be
a primary factor determining the observed properties of ultracompact jets.
We discuss possible applications of opacity effects
to studying the conditions existing in central regions of active galactic nuclei.


\end{abstract}

\section {Introduction}

Extragalactic jets are believed to be formed in the vicinity of an accretion disk
around a supermassive black hole in the center of active galactic nuclei (AGN). 
Pressure and density 
gradients, as well as a toroidally shaped magnetic field produced by the 
rotation of the disk, may collimate the 
outflowing material into two bipolar streams, and accelerate it
up to relativistic speeds (see reviews by Begelman,
Blandford, \& Rees, 1984; Begelman 1995).
Physical conditions in the immediate proximity of an accretion disk 
determine many aspects of jet evolution at
larger spatial scales. The extent of this {\it ultracompact} fraction of 
the jet is estimated to be $\sim0.1-10$\,pc. 
Roland et al. (1994) model it as a turbulent 
region of $\lea 1$\,pc in size. Marscher \& Gear (1985) estimate the 
projected size of ultracompact jet to be $\sim 0.3h^{-1}$\,pc, 
based on VLBI observations of a radio flare in 3C\,273. 
K\"onigl (1981) and Zensus,
Cohen, \& Unwin (1995) regard an unresolved core 
as an ultracompact jet. 

In images obtained with Very Long Baseline Interferometry (VLBI), the core is 
usually identified with the most compact (often unresolved)
feature exhibiting a substantial flux and flat spectrum across the radio band.
 At any given frequency,
the core is believed to be located in the region of the jet where the optical 
depth is $\tau=1$.
The core absolute position, $r_{\rm core}$, should therefore depend on the observing
frequency, $\nu_{\rm obs}$. K\"onigl (1981) gives $r_{\rm core} \propto 
(\nu_{\rm obs})^{-1/k_{\rm r}}$. The power index $k_{\rm r}$ depends on the shape of
electron energy spectrum and on the magnetic field and particle density 
distributions in the ultracompact jet.
Observed offsets of the core position at different frequencies have been 
reported
for several sources including 1038+528\,A (Marcaide $\&$ Shapiro 1984), 
4C\,39.25 (Guirado et al. 1995), 3C\,395 
(Lara et al. 1996), and 3C\,309.1 (Aaron 1996).
Several studies of the core position offset have been undertaken for
3C\,345
(Biretta, Moore, \& Cohen. 1986; Unwin et al. 1994; 
Zensus et al. 1995).

In this paper, we discuss synchrotron self--absorption and free--free absorption  
in the nuclear regions of AGN. We use the frequency dependence of the
VLBI core position
as a tool for determining the physical conditions of 
ultracompact  jets. In section \ref{sc:theory}, we describe 
a model adopted for ultracompact jets, and outline the relations between
core shift and physical properties of the jets. Measurements of the shift of VLBI
core in radio sources are discussed in section 
\ref{sc:obs_shift}.
In section \ref{sc:jets}, the measured core offsets are applied to deriving \
the magnetic field distribution and 
physical conditions in the central regions of Cygnus A, 3C\,309.1, 3C\,345,
3C\,395, 4C\,39.25, and 1038+528\,A.

Throughout the paper, we use a Hubble constant $H_0 = 100\,h$\,km\,s$^{-1}$\,Mpc$^{-1}$
and deceleration parameter $q_0 = 0.5$.
Unless defined otherwise, all quantities are in the {\it cgs} units, except for
distances which are given in parsecs.

\section{Ultracompact jets \label{sc:theory}}

A region of significant particle acceleration is assumed to exist at the very
center of an AGN. 
 The accelerated particles are confined by an ambient
plasma with steep pressure and density gradients along the rotational axis of
the central engine. A highly collimated  relativistic outflow (jet) can be formed 
in result.
The jet can contain both protons and electron--positron pairs (Sol et al. 1989).
 Interactions between the jet and the ambient
medium are usually ignored, and the jet itself is assumed to have no
transverse velocity gradient. The jet hydrodynamics is then given by a stationary
adiabatic flow satisfying Bernoulli's equation $({\bf v \cdot \nabla}) \gamma_{\rm j}
p^{1/4}=0$ \cite{dm88}.  
A nozzle is formed at a distance $r_{\rm c}$ where the pressure $p$ drops below 4/9 of
its initial value, and the flow becomes supersonic, with bulk Lorentz factor
$\gamma_{\rm j}(r_{\rm c}) = 1.22$. Beyond the
nozzle, the flow is accelerated by the conversion of internal relativistic 
particle energy $\gamma_{\rm e}$ to bulk kinetic energy $\gamma_{\rm kin}$, so that $\gamma_{\rm kin}
\propto \gamma_{\rm e} \propto r^{\epsilon}$ ($\epsilon=$const). The resulting particle
energy distribution $N(\gamma_{\rm e})$ can be approximated by a power law:
$N(\gamma_{\rm e}) = N_0 \gamma_{\rm e}^{-s}$ for $\gamma_{\rm min}(r)<\gamma_{\rm e}<\gamma_{\rm max}(r)$,
where $s$ is the particle energy spectral index. The jet radiation is then described 
by the inhomogeneous synchrotron spectrum with spectral index $\alpha = (1-s)/2$.
Unless otherwise specified, the jet bulk Lorentz factor $\gamma_{\rm j}$, and viewing
angle $\theta$ are assumed constant. 
The jet geometry can be approximated by a conical or paraboloidal expansion,
with the transverse radius 
$R=a\,r^{\varepsilon}$ ($\varepsilon\le 1$), for $r_{\rm c}\le r\le r_{\rm max}$,
where $r_{\rm max}$ can be taken $\sim 100$--$1000\,r_{\rm c}$, for most of parsec-scale jets.
The jet opening angle $\phi = \arctan(R/r)$ is constant for
$\varepsilon=1$.
The magnetic field and particle density decrease with $r$, and can be
approximated as: $B=B_1(r_1/r)^{m}$ and $N=N_1(r_1/r)^{n}$,
where $B_1$, $N_1$ are the magnetic field and the electron density at
$r_1 = 1$\,pc. K\"onigl (1981) shows that the combination  $m=1$, $n=2$
can be used to describe the observed X-ray and synchrotron emission from 
the most compact regions of VLBI jets.

\subsection{VLBI core and ultracompact jet}

In the scheme described above, the protons become subrelativistic in
the rest frame of the flow outside a characteristic radius $r_0$, and 
the bulk Lorentz factor freezes at the value $\sim\gamma_{\rm j} =
\gamma(r_0)$. This location in the jet is most likely to be observed
as the VLBI core \cite{mar95}. It is also referred to as the
``injection point'' (at which the relativistic plasma is assumed to be
injected into the jet at $\gamma_{\rm j}$) in most of the models dealing
with parsec-scale relativistic jets. 

At any given frequency, the VLBI core is observed at the location where
the optical depth of synchrotron self--absorption is $\tau_{\rm s}=1$. 
For given $N(r)$ and $B(r)$, the corresponding $\tau_{\rm s}$ is 
(e.g. Rybicki \& Lightman 1979):
\begin{equation}
\label{eq:theor0}
\tau_{\rm s}(r) = C(\alpha)N_1 \left(\frac{e B_1}{2\pi m_{\rm e}}\right)^\epsilon
\frac{\delta^\epsilon \phi_o}{r^{(\epsilon m + n - 1)} \nu^{\epsilon+1}}\,.
\end{equation}
Here $m_{\rm e}$ is the electron mass and $\epsilon = 3/2-\alpha$.
The observed jet opening angle $\phi_o = \phi\csc\theta$.
$C_2(\alpha)$ is described in Blumenthal \& Gould (1970), and  
$C_2(-0.5)=8.4\cdot 10^{10}$. 
Setting $\tau_{\rm s}(r)$ to unity gives for the distance 
from the central engine to the core observed at frequency $\nu$ 
\begin{equation}
\label{eq:theor1}
r[{\rm pc}] = (B_1^{k_{\rm b}} F/\nu)^{1/k_{\rm r}}\,,
\end{equation}
with
\begin{equation}
\label{eq:theor2}
F = (1+z)^{-1}[6.2\cdot 10^{18} C_2(\alpha)\delta_{\rm j}^{\epsilon}N_1 \phi_o]^{1/(\epsilon+1)}\,.
\end{equation}
From (\ref{eq:theor0}), $k_{\rm r} = ((3-2\alpha)m
+2n-2)/(5-2\alpha)$ and $k_{\rm b} = (3-2\alpha)/(5-2\alpha)$.  For the equipartition
value of $k_{\rm r} = 1$, the choice of $m=1$, $n=2$ appears to be the 
most reasonable.
The corresponding $k_{\rm r}$ in this case does not depend on spectral index,
and both $B(r)$ and $N(r)$ decline (taking $m=2$, for instance, would
result in $n=0.5+\alpha$, which is unrealistic: $N(r)$ would then remain 
nearly constant, or even grow, with increasing $r$). 

If the apparent core positions measured at two frequencies 
$\nu_1,\nu_2\,(\nu_1<\nu_2)$ 
differ by $\Delta r_{\rm mas}$, one can introduce a measure of core position
offset
\begin{equation}
\label{eq:theor3}
\Omega_{r\nu} = 4.85\cdot 10^{-9}\frac{\Delta r_{\rm mas} D_{\rm L}}{(1+z)^2} \cdot 
\frac{\nu_1^{1/k_{\rm r}} \nu_2^{1/k_{\rm r}}}{\nu_2^{1/k_{\rm r}}-\nu_1^{1/k_{\rm r}}} \,,
\end{equation}
where $D_{\rm L}$ is the luminosity distance to the source, and frequencies are
given in Hz. $\Omega_{r\nu}$ can be used for assessing the quality of core 
position measurements. For ideal measurements, 
$\Omega_{r\nu} = (B_1^{k_{\rm b}}F)^{1/k_{\rm r}} \sin\theta = {\rm const}$, for all
frequency pairs. A dispersion in the values of $\Omega_{r\nu}$ derived from
different frequency pairs reflects the inaccuracy of core position data. 
If $k_{\rm r}$ has been determined inaccurately, or if it varies along the jet, 
then $\Omega_{r\nu}$ will exhibit a systematic trend.
The variations of $k_{\rm r}$ can occur when
the jet crosses a region with rapidly changing absorption properties
(e.g. the broad line region). In this case, 
$\Omega_{r\nu}$ can be used to estimate the change of $k_{\rm r}$. If two values
$\Omega_{r\nu 1,2}$ (measured between frequencies $\nu_1$ and $\nu_2$) and
$\Omega_{r\nu 2,3}$ (between $\nu_2$ and $\nu_3$) are different, the relation
between the corresponding $k_{\rm r}$'s is:
\begin{equation}
\label{eq:theor3a}
k_{r2,3} \approx k_{r1,2} \left[\log(\Omega_{r\nu 1,2})/\log(\Omega_{r\nu 1,3})\right]\,.
\end{equation}
With a measured $\Omega_{r\nu}$, equation (\ref{eq:theor1}) can be applied to
determine the magnetic 
field in the jet at $r=1$\,pc
\begin{equation}
\label{eq:theor4}
B_1 = (\Omega_{r\nu}/\sin\theta)^{k_{\rm r}/k_{\rm b}} F^{-1/k_{\rm b}}\,,
\end{equation}
and the absolute distance of the core from the central engine:
\begin{equation}
\label{eq:theor5}
r_{\rm core}(\nu) = \Omega_{r\nu}\left[\nu^{1/k_{\rm r}}\sin\theta\right]^{-1}.
\end{equation}
Formally, $r_{\rm core}$ refers to the distance between the core and the sonic point,
$r_{\rm c}$. However, since $r_{\rm core} \gg r_{\rm c}$, in most cases, $r_{\rm core}$ can be taken as
a fair approximation for the distance to central engine.
It should also be stressed that $r_{\rm core}$ in (\ref{eq:theor5}) depends only on 
measured quantities ($\Delta r_{\rm mas}$ and $k_{\rm r}$) and on the jet viewing
angle $\theta$ that can be determined from observations.
Therefore $r_{\rm core}$ can be used as a reliable estimate of core position, 
or at least of the projected distance from the central engine (if $\theta$
is poorly known).
From (\ref{eq:theor4}) and (\ref{eq:theor5}), the magnetic field in the
VLBI core
observed at frequency $\nu$ is given by
\begin{equation}
\label{eq:theor6}
B_{\rm core}(\nu)\approx \nu^{m/k_{\rm r}} \left[\frac{\Omega_{r\nu}}{(1+z) 
\sin\theta  } \right]^{\zeta}
F^{-1/k_{\rm b}} \,,
\end{equation}
where $\zeta = (k_{\rm r}/k_{\rm b}) - m$. 
For a typical $\alpha=-0.5$, we have $\zeta=0.5$. 

\subsection{Equipartition regime}

We now consider the equipartition between jet particle and magnetic field energy 
densities, with the magnetic field energy density given 
by $k_{\rm e}\Lambda B^2 /\pi$ \cite{bk79}, where $k_{\rm e} \lea 1$, 
$\Lambda = \ln 
(\gamma_{\rm max}/ \gamma_{\rm min})$. In this case, $k_{\rm r}=1$ (with $m=1$, $n=2$), and
the core position offsets can also be used to determine the 
total (kinetic + magnetic field) power of the jet 
\begin{equation}
\label{eq:theor7}
L_t\approx 2.03\cdot 10^{29} \left[\Omega_{r\nu} (1+z) \phi_o\right]^{3/2}
\frac{\Theta(3 + 2k_{\rm e}\Lambda) \gamma_{\rm j}^2 \beta_{\rm j}}
{k_{\rm e}^{1/2} \delta_{\rm j} \sin\theta}\,,
\end{equation}
where $\Theta = \ln(r_{\rm max}/r_{\rm c})$. 
The $L_t$ can be further used to derive the magnetic field at 1\,pc
\begin{equation}
\label{eq:theor8}
B_1 \approx 2.92\cdot 10^{-9} \left[\frac{\Omega_{r\nu}^3(1+z)^3}
{k_{\rm e} \delta_{\rm j}^2 \phi \sin^5\theta}\right]^{1/4}\,.
\end{equation}
The magnetic field in the core can be obtained, similarly to  
$B_{\rm core}$ derived in (\ref{eq:theor6}). The equipartition can
also be used for predicting the core offset in a source with known
synchrotron luminosity $L_{\rm syn}$. In this case, the expected shift 
of the core position between two frequencies $\nu_1$ and $\nu_2$ is
\begin{equation}
\label{eq:theor9}
\Delta r_{\rm mas} \approx \frac{C_{\rm r}(1+z)}{D_{\rm L}\gamma^2
\phi_o}\frac{\nu_2 - \nu_1}{\nu_1 \nu_2}
\left[ \frac{L_{\rm syn}\sin\theta}
{\beta(1-\beta\cos\theta) \Theta} \right]^{2/3}\,,
\end{equation}
with $C_{\rm r} = 4.56\cdot 10^{-12}$, and assuming $k_{\rm e}=1$. Table \ref{tb:shifts} contains first order
predictions of the core shift between 5 and 22\,GHz, for several prominent 
radio sources. 

\begin{table}[t]
\caption{Predicted core position shift between 5 and 22\,GHz \label{tb:shifts}}
\footnotesize
\begin{center}
\medskip
\begin{tabular}{rccc} \hline\hline
Name & Type  & $\Delta r_{5,22}$ & $L_{\rm syn}$ \\ 
 & & [mas] & [erg\,s$^{-1}$] \\\hline
3C\,273 & LPQ & 0.78 & $2.4\cdot 10^{43}$ \\
3C\,216 & HPQ & 0.70 & $3.1\cdot 10^{43}$ \\
3C\,120 & LPQ & 0.61 & $8.4\cdot 10^{41}$ \\
4C\,39.25  & LPQ & 0.57 & $2.2\cdot 10^{44}$ \\
3C\,345 & HPQ & 0.37 & $1.2\cdot 10^{44}$ \\
1807+698 & BL & 0.26 & $3.3\cdot 10^{41}$ \\
BL Lac & BL & 0.11 & $2.3\cdot 10^{41}$ \\ \hline
\end{tabular}
\end{center}
\end{table}

The 
synchrotron luminosities in Table
\ref{tb:shifts} are calculated from a database compiled by Ghisellini 
et al. (1992), using the model of Blandford \& K\"onigl (1979) with
$\gamma_{\rm max}/\gamma_{\rm min}=100$, $r_{\rm max}/r_{\rm c}=100$. The jet opening
angle $\phi=0.5\deg$ was used for all sources. The synchrotron 
self--Compton Doppler factors from the database are used  for estimating 
the jet viewing angles and Lorentz factors, assuming the minimum jet 
kinetic power condition $\gamma_{\rm min}=1/\sin\theta$. This gives lower limits
for synchrotron luminosities, and upper limits for the shifts. One can see,
from Table \ref{tb:shifts}, that the core shifts may be noticeably large, 
and can be measured from VLBI data.

\subsection{External pressure gradients \label{sc:pressure}}

\begin{figure}[t]
\cl{\psfig{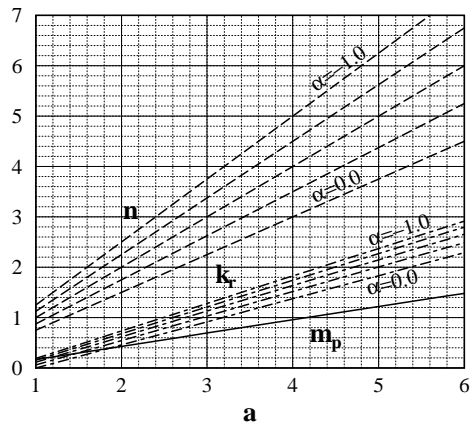}}
\caption{Physical conditions in the jet in the presence of strong
pressure gradients. Power index $a$ describes the pressure gradient
($p\propto r^{-a}$) along the jet axis.  The corresponding magnetic
field and particle density gradients have power indices $m_{\rm p}$
(for $m=1$) and $n$, respectively.  The $n$ and $k_{\rm r}$ are given
for different values of synchrotron spectral index
$\alpha$. \label{fg:gradients}}
\end{figure}

In the vicinity of an accretion disk, physical conditions in the jet become
sensitive to the gradients of the pressure $p\propto r^{-a}$ of the 
ambient medium. Since $\gamma_{\rm j} p^{1/4} = const$, to satisfy Bernoulli's equation,
the jet Lorentz factor and opening angle vary. Consequently, the corresponding
gradients in $B(r)$ and $N(r)$ should also change along the jet in this region \cite{gm96}. 
The dependences of magnetic field and particle density distributions on pressure
gradients are shown in Fig. \ref{fg:gradients}. If $k_{\rm r}$ and spectral index $\alpha$
are measured, one can find, from Fig. \ref{fg:gradients}, the particle density and 
magnetic field distribution satisfying the conditions of stationary adiabatic
flow. The magnetic field at $r\gg r_{\rm c}$ is $\propto r^{-m_{\rm p}}$, $m_{\rm p} = m\,a/4$. The 
particle density distribution can be approximated by $n = a(3-2\alpha)/4$.

\begin{figure}[t]
\cl{\psfig{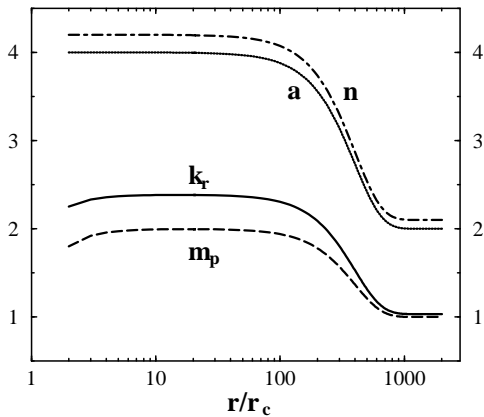}}
\caption{Change of physical conditions along the jet axis, $r$. The
pressure gradient $a$ corresponds to gaseous clouds supported by
thermal pressure and maintaining a mass distribution with spherically
symmetrical gravitational potential. The cloud region extends up to
400$r_{\rm c}$ ($r_{\rm c}$ refers to the distance at which the jet
becomes supersonic). The equipartition regime is approached at the
outer boundary of the cloud region, with $m_{\rm p}=1$, $n=2$, $k_{\rm
r}=1$. Significant deviations from the equipartition are seen on
smaller scales, resulting in stronger self--absorption in the inner
parts of the jet ($k_{\rm r} \le 2.5$). \label{fg:blrmod}}
\end{figure}

If the jet is confined by gaseous clouds supported by thermal pressure from the
central source and the stellar population with star density $\rho_{\rm s}$, the  ambient
medium density can be modelled by an exponential decrease $p(r) = p_0 \exp(-r^2/d^2)$,
with the characteristic size of the cloud system
 $d = (3kT/\pi G \rho_{\rm s} m_{\rm p})^{0.5}$ \cite{br74}. 
Here $m_{\rm p}$ is the protom mass, and $T$ is the ambient plasma temperature.
Such a distribution reflects the likely conditions in the broad line region (BLR)
surrounding the nucleus of an AGN.
The jet properties in this case are shown in Fig. \ref{fg:blrmod}, for
$d = 400 r_{\rm c}$, $a(r_{\rm c}) = 4$, and $m=2$. The parameters are chosen so as to
approach the equipartition at $r\approx3d$. The jet opacity to synchrotron 
self--absorption increases significantly in the region of steepest pressure
gradients. This implies that confinement effects may influence the observed
frequency dependence of the VLBI core position in some radio sources. If $\rho_{\rm s}$ and 
$T$ are known, the core position measurements can be used 
for estimating the size of nuclear cloud system.

\subsection{Free--free absorption \label{sc:absorption}}

Foreground free--free absorption in a hydrogen plasma is expected to affect 
the VLBI--scale radio emission propagating through a dense nuclear environment.
The existence of absorbing, 
circumnuclear plasma has been suggested in 3C\,84 (Vermeulen et al. 1994; 
Walker et al. 1994), Cen\,\,A (Jones et al. 1996), and Cyg\,\,A (Krichbaum et al.
1997). In all three sources, only the counter--jet emission is believed to be absorbed. 
One can also expect to find absorption in the jet--side emission, if the jet viewing 
angle is relatively large, and the absorbing medium extends sufficiently
high above the accretion disk plane. One possibility would be the broad line 
region formed by hydrogen plasma clouds entrained from the disk
\cite{cr93}.
The optical depth of free--free absorption is given by (e.g. Levinson 
et al. 1995)
\begin{equation}
\label{eq:theor11}
\tau_{\rm f}(r) = 5\cdot 10^{16} T^{-1.5} \nu^{-2} \bar{g}  n_{\rm e}(r) n_{\rm i}(r) l_{\rm pc}\,,
\end{equation}
where $l_{\rm pc}$ is the size of absorbing region. Assuming a pure hydrogen plasma with 
uniform  density, we
can take $n_{\rm e} \approx n_{\rm i}$. The hydrogenic free--free Gaunt 
factor, $\bar{g}$, can be evaluated numerically (Hummer 1988). 
An analytical long--wave approximation of $\bar{g}$ given by 
Scheuer (1960) can also be used.

The typical sizes of BLR inferred from reverberation mapping and modelling the
optical broad line emission of AGN vary between 0.01 and 1\,pc (e.g. Kaspi et al.
1996; Baldwin et al. 1996). The corresponding densities of hydrogen plasma are
$n_{\rm H} \sim 10^8-10^{12}$\,cm$^{-3}$. The disk--wind model \cite{cr93} predicts the 
BLR sizes of up to $\sim 20$\,pc, with $n_{\rm H}$ between $10^6$
and $10^{12}$\,cm$^{-3}$. There is evidence that weaker emission lines may originate
from a very extended ($r\le400$\,pc) thermal component with densities $n_{\rm H} \sim
10^2-10^8$\,cm$^{-3}$ and electron temperatures $T_{\rm e}\sim10^4-10^6$\,K \cite{fkf97}.
Most of these plasmas can alter significantly the GHz--range emission from
VLBI jets (for instance, $\tau_{\rm f} =250$ at 10\,GHz, for $n_{\rm e}=10^{6}$\,cm$^{-3}$,
$T=10^4$\,K, $l_{\rm pc}=0.1$). Udomprasert et al. (1997) report very high rotation
measures, $RM\approx 40\,000$\,rad\,m$^{-2}$ in the VLBI core of the quasar OQ\,172, which further 
supports the presence of high--density thermal medium around the ultracompact
jets.

For a spherical distribution of BLR clouds, we can take 
$n_{\rm e}(r) \propto \varepsilon n_0 r^{-n}$
($\varepsilon$ is the volume filling factor of the cloud distribution).
Then, from (\ref{eq:theor11}), $\tau_{\rm f} \propto r^{-2n+1}$.
A crude estimate $n\approx 3$ can be adopted (remembering that $n$
refers essentially to plasma density variations along the jet axis).
Then $r_{\rm core} \propto \nu^{-2.5}$, and the corresponding core shift is roughly
10 times smaller than that due to synchrotron self--absorption at typical VLBI 
observing frequencies. 

\section {Measuring the core position offsets \label{sc:obs_shift}}

High-precision measurements of radio source absolute positions which are
required for alignment of VLBI images at different frequencies are not 
readily available in most cases.
Extensive absolute astrometry
(see Fomalont 1995) observations are necessary, in order to establish 
a reliable link between 
VLBI data at different frequencies. In quasi--simultaneous 
multifrequency  observations, the phase--referencing technique 
\cite{bc95} can be sufficient
for the purpose of image alignment. If neither of the abovementioned techniques
is available, the frequency dependent shift of the core position can be 
deduced from
comparison of observations made at close epochs, assuming that the
moving features observed in parsec--scale jets are optically thin and
therefore should have their positions unchanged. 
In this case, the offsets between the component locations measured at different 
frequencies will reflect the frequency dependent shift of the position of the
source core. 

\subsection{Core shift in 3C\,345 \label{sc:345shift}}

An extensive long--term VLBI monitoring database is available for 3C\,345
(Zensus et al. 1995, 1997; Lobanov 1996).
In the data for 3C\,345, there are three close pairs of VLBI observations:
(1989.24 at 22.2\,GHz)--(1989.26 at 10.6\,GHz), 
(1992.44 at 22.2\,GHz)--(1992.45 at 5\,GHz), 
(1993.69 at 5\,GHz)--(1993.72 at 22.2\,GHz). In these pairs, the separations 
between the observations in these pairs do not exceed 10 days.
We also use two other
close pairs with separations of 51 days: (1992.71 at 8.4\,GHz)--(1992.86 at
22.2\,GHz), and 70 days: (1993.69 at 5\,GHz)--(1993.88 at 8.4\,GHz).
We measure the offsets in the closest and brightest components
which have most reliably measured positions. For the pairs
with 51 and 70 days separation, the positions of jet components have been 
corrected for the proper motions measured from the polynomial fits to the 
components trajectories (Lobanov 1996).

\begin{table}[t]
\caption{Average offsets of the core position in 3C\,345\label{tb:offset}}
\begin{center}
\begin{tabular}{rcccc} \hline\hline
 1 & 2& 3& 4& 5\\ \hline
Frequency pair & $\Delta r$ & $b_{\Delta r}$ &
$\Delta \phi$ & $\Delta r_{\rm proj}$ \\ 
& [mas] &  & [deg] & [pc] \\ \hline\hline
5.0--~8.4\,GHz  & $0.05\pm0.03$ & 5\%  & $-70.8\pm4.1$ & 0.1 \\
10.6--22.2\,GHz & $0.17\pm0.05$ & 20\% & $-73.5\pm2.0$ & 0.6 \\
8.4--22.2\,GHz  & $0.21\pm0.06$ & 20\% & $-71.4\pm1.4$ & 0.8 \\
5.0--22.2\,GHz  & $0.33\pm0.10$ & 28\% & $-77.1\pm2.6$ & 1.2 \\
\hline
\end{tabular}
\medskip
\end{center}
Notes: 2 -- offset radius in mas; 3 -- corresponding fraction of restoring beam; 
4 -- position angle of the offset direction; 5 -- projected linear distance corresponding to 
the offset.
\end{table}

\begin{figure}[t]
\cl{\psfig{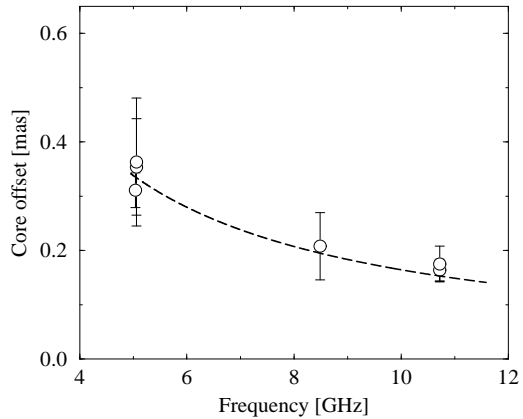}}
\caption{Frequency dependent shift of the core position in
3C\,345. Dashed line corresponds to $r\propto \nu^{-1.04}$. Reference 
frequency is 22.2\,GHz.\label{fg:offset}}
\end{figure}

The average angular offsets are compared in Table \ref{tb:offset} with the
restoring beam sizes of spectral index maps
made from the corresponding frequency pairs. 
The direction of the core shift ($~-75\deg$) is similar to the position angle of the innermost
part of the jet observed at 43\,GHz (Krichbaum \& Witzel 1992). 
Between 5 and 8.4\,GHz, the position shift cannot be measured reliably because of
the insufficient resolution and errors due to the proper motion corrections. 
The measured shift constitutes only about 5\% of  the 
restoring beam.
With increasing frequency separation, the position shift becomes a 
prominent fraction of beamsize (up to 27\% for spectral index maps 
between 5 and 22.2\,GHz). 
Fig. \ref{fg:offset} shows the measured offsets with respect to the reference frequency
(22.2\,GHz). The 5--8.4\,GHz measurements are not included.
The dashed line in Fig. \ref{fg:offset} represents the best fit:
$r \propto \nu^{(-1.04\pm 0.16)}$.

\begin{table*}
\caption{Model and derived parameters of ultracompact jets\label{tb:jet-model}}
\footnotesize
\begin{center}
\medskip
\begin{tabular}{rccccccc} \hline\hline
\multicolumn{8}{c}{\sc A. Model parameters} \\\hline
Name & $z$ & $\gamma_{\rm j}$ & $\theta_{\rm j}$ & 
$\phi_{\rm j}$ & $\alpha_{\rm j}$ & $N_1$ & Refs. \\
      &        &  & [deg] & [deg] &  & [cm$^{-3}$] & \\\hline
Cyg A & 0.0562 & 5.0 & 80.0 & 7.0 & $-$0.5 & 8000 &  1,2\\ 
3C309.1 & 0.905 & 5.0 & 20.0 & 2.0 & $-0.6$ & 17000$^{\dagger}$ & 3  \\
3C345   & 0.594 & 5.6 & 7.6 & 2.4 & $-$0.6 & 1300$^{\dagger}$ & 4,5\\
3C395   & 0.635 & 7.1 & 8.7  & 0.5 & $-0.6$ & 1800$^{\dagger}$ & 6 \\
4C39.25 & 0.699 & 11.0 & 30.0 & 4.0  & $-0.5$ & 1000$^{\dagger}$ & 7 \\
1038+528A& 0.678 & 4.0 & 20.0 & 5.0 & $-0.6$ & 1900$^{\dagger}$ & 8 \\\hline
\end{tabular}
\end{center}
References: 1 -- Krichbaum et al. 1997; 2 -- Carilli \& Barthel 1996;
3 -- Kus 1993; 4 -- Zensus et al. 1997; 5 -- Lobanov 1996; 6 -- Lara et al. 1994; 
7 -- Alberdi et al. 1993; 8 -- Marcaide et al. 1994.
\end{table*}
\begin{table*}
\footnotesize
\begin{center}
\medskip
\begin{tabular}{rcccccccc} \hline\hline
\multicolumn{9}{c}{\sc B. Measured and derived parameters$^{\S}$ } \\\hline
Name & $\nu_1/\nu_2$ &  $\Omega_{r\nu}$ & $B_1$ & $B_{\rm core}^{\dagger}$ & $r_{\rm core}^{\dagger}$ & $L_{\rm tot}$ & $L_{\rm syn}$ & $T_{\rm b\,max}$ \\
     & [GHz] & [pc\,GHz$^{-1}$] & [G]   & [G]  &  [pc] & [10$^{46}$\,erg\,s$^{-1}$] & [10$^{46}$\,erg\,s$^{-1}$] & [10$^{11}$K] \\ \hline\hline
Cyg A  & 22/43 & $2.1\pm0.4$ & $0.07\pm0.01$ & $0.7\pm0.3$ & $0.10\pm0.02$ & $0.55\pm0.05$ & $0.07\pm0.01$ & $1.0$ \\ \hline
3C309.1 & 1.5/2.3  & $16.3\pm7.2$ & $1.2\pm0.4$& $0.6\pm0.4$& $2.1\pm0.9$& $5.8\pm1.7$&
$0.71\pm0.21$& $7.2$ \\
        & 1.6/2.3  & $15.4\pm8.8$ & $1.1\pm0.5$& $0.5\pm0.4$& $2.0\pm1.2$& $5.3\pm2.3$&
$0.66\pm0.28$& $7.2$ \\
        & 2.3/5.0  & $7.2\pm3.6$& $0.4\pm0.1$& $0.4\pm0.3$& $0.9\pm0.5$& $1.7\pm0.6$&
$0.21\pm0.07$& $6.5$ \\
        & 5.0/8.4      & $<5.2$& $<0.2$& $<0.3$& $<0.7$& $<1.0$& $<0.1$& $<6.3$  \\
        & 8.4/15.1   & $<8.0$& $<0.4$& $<0.4$& $<1.0$& $<2.0$& $<0.2$& $<6.6$ \\\hline
3C345   & 10.6/22.2 & $13.1\pm3.8$ & $1.9\pm0.3$ & $0.4\pm0.2$ & $4.4\pm1.3$ & $19.8\pm3.2$&
$2.4\pm0.4$ & $25.8$ \\
        & 8.4/22.2 & $10.8\pm3.1$ & $1.4\pm0.6$ & $0.4\pm0.2$ & $3.7\pm1.0$ & $14.7\pm2.2$ &
$1.8\pm0.3$ & $25.1$   \\
        & 5.0/22.2 & $8.1\pm2.4$ & $0.9\pm0.2$& $0.3\pm0.2$& $2.8\pm0.8$& $9.6\pm1.6$&
$1.2\pm0.2$& $24.2$ \\ \hline
3C395   & 2.3/8.4  & $7.4\pm2.4$& $1.4\pm0.6$& $0.7\pm0.4$& $2.2\pm0.7$& $1.0\pm0.2$&
$0.12\pm0.02$& 22.2 \\ \hline
4C39.25 & 2.3/8.4& $6.3\pm1.3$& $0.6\pm0.2$& $1.0\pm0.4$& $0.6\pm0.1$& $18.2\pm1.6$&
$2.2\pm0.2$& 2.4 \\ \hline
1038+528A& 2.3/8.4 & $6.3\pm1.2$& $0.4\pm0.1$& $0.5\pm0.2$& $0.8\pm0.2$& $2.6\pm0.2$&
$0.32\pm0.03$& 7.4 \\
& 8.4/15.1 & $<1.5$& $<0.13$& $<0.3$& $<0.2$&  $<0.3$& $<0.04$& $<6.2$ \\\hline
\end{tabular}
\end{center}
Notes: $^{\S}$ -- only the errors due to uncertainties of the core offset measurements
are considered; $\dagger$ -- derived by equating $B_1$ to its equipartition value;
$\ddagger$ -- $B_{\rm core}$ and $r_{\rm core}$ are given for $\nu_{\rm obs}=22.2$\,GHz.
\end{table*}

\subsection{Effect of the reference point offset on spectral imaging 
\label{sc:refpoint}}

\begin{figure}[t]
\cl{\psfig{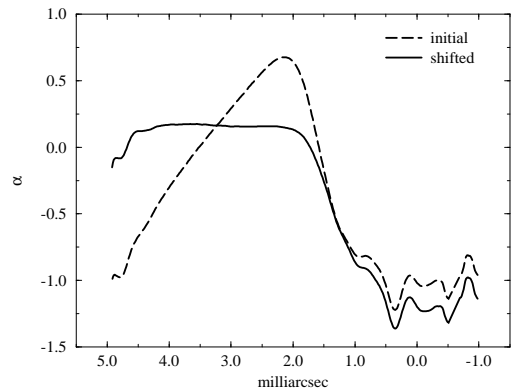}}
\caption{Spectral index profiles taken in the nuclear region of 3C345 (5--22.2\,GHz, 
September 93). Dashed line is the spectral index profile obtained by aligning both
images by their respective core positions. Solid line is the profile obtained after
applying the measured position offset of the core.\label{fg:spmap}}
\end{figure}

Whenever an offset constitutes a significant fraction of restoring
beam, it can influence substantially the derived spectral
properties. An example of this effect is shown in Fig.
\ref{fg:spmap}. 
The 5--22\,GHz spectral index profiles shown in Fig. \ref{fg:spmap} 
are measured along the jet ridge line in the nuclear
region of 3C\,345. The core extends from $+2$ to $+5$\,mas.
The profile obtained by aligning the images on their
respective core positions  
shows a strong gradient, with optically thin spectral index across a
significant fraction of the core. Applying the measured core position offset
between 5 and 22.2\,GHz levels the spectral index across the entire core, with
$\alpha_{\rm core}\approx 0.2$.
Similar
corrections applied to several other spectral index maps of 3C\,345
have resulted in decreased peak values, and smoother spectral index 
distributions across the nuclear region.

\section{Properties of ultracompact jets\label{sc:jets}}

To illustrate the implications of frequency dependence of VLBI core position,
we calculate physical properties of ultracompact jets in 3C\,345 and 5 other AGN 
with reported core position offsets. Table \ref{tb:jet-model} summarizes
the model parameters (panel A), and gives the measured and derived quantities (panel
B). For all sources, we assume $m=1$, $n=2$, and use crude estimates 
$\gamma_{\rm max}/\gamma_{\rm min} = 100$, and 
$r_{\rm max}/r_{\rm c} = 100$ (since only the logarithms of these ratios affect the
calculated luminosities). The value of $N_1$ is determined in most cases, by setting 
$B_1$ to its equipartition value given by (\ref{eq:theor8}). Synchrotron luminosity,
$L_{\rm syn}$, and maximum brightness temperature, $T_{\rm b\,max}$, are calculated using
the model of Blandford and K\"onigl (1979). For the purpose of
comparison between different sources, the offset measures,
$\Omega_{r\nu}$ are given, rather than the angular offsets cited in the original
publications.

\subsection{Cygnus A \label{sc:cygnus}}

We adopt the component identification made by Krichbaum et al. (1997), and use their
 position measurements to derive the offset of the core in Cygnus A between 22.2 and\
43.2\,GHz (observation epochs 1992.44 and 1990.40, respectively). Krichbaum et al.
(1997) give Gaussian model fits describing the compact structure of the source. 
We use these models to calculate the
shifts of the total of four features situated both in the jet (components J3, J4) 
and counter--jet (components C1, C2). The two closest jet--side components, J1 and J2,
are blended together in the model at 22\,GHz. The averaged shifts are:  
$\Delta r_{\rm j} = 74\,\mu$as for J3 and J4, and $\Delta r_{\rm cj} = -54\,\mu$as for
C1 and C2. The negative sign for the shift on the counter--jet side indicates that
the component separations from the core are shorter at 43.2\,GHz. 
This fact further
supports the identification, as well as frequency dependence, of the core location in
Cygnus A. We then estimate that the shift of the core position amounts to 
$64\pm12\,\mu$as, and use this value for calculating the physical properties of
the jet. 

The estimated $B_1=0.07$\,G is consistent with the field strength ($B_{\rm 1\,pc}\approx
0.06$\,G, for $\phi_{\rm j} = 7.0\deg$) in a Poynting
flux jet \cite{lr96} driven by a $3\cdot10^8$\,M$\solar$ black hole. Carilli \& Barthel (1996,
CR96 hereafter) use
the unresolved core flux density of Cygnus A at 43\,GHz, and apply the minimum 
energy equation, to arrive at $B_{\rm m}=0.16$\,G in a nuclear region of 0.15\,mas 
(0.11\,pc) in size. We have
$r_{\rm 43\,GHz} \approx 0.05$\,pc, for the distance from the jet origin to
the core observed at 43\,GHz. With these values, 
$B_1 = 0.11 B_{\rm m}\tan\phi_{\rm j}/r_{\rm 43\,GHz} \approx 0.06$\,G.

The total jet power, $L_{\rm tot}$, can be related to the luminosity of
extended radio lobe, $L_{\rm R}$: $L_{\rm tot} = L_{\rm R}/\eta$ (CR96), with $\eta=$0.01--0.1 
(Leahy 1991) describing the efficiency of converting bulk kinetic energy into radio 
luminosity. In Cygnus A, the lobe radio luminosity 
$L_{\rm R} = 4.4\cdot 10^{44}$\,erg\,s$^{-1}$ (CR96), which results in $\eta=0.08$, for
our derived $L_{\rm tot}$.

The predicted location, $r_{\rm c.e.}$, of the central engine in Cygnus A is offset 
by about 0.14\,mas
from the core. Comparison between the 22\,GHz flux densities of two bright features 
(J2 on the jet side, and C1 on the counter--jet side) evenly separated from 
$r_{\rm c.e.}$ gives $2.2\pm0.3$ for the jet--to--counter--jet ratio (this figure should
be viewed as an upper limit, if a foregorund absorbing medium is present). The calculated 
ratio results
in $\theta_{\rm j}= 80.6\pm 1.2$ for the adopted $\gamma_{\rm j}=5$. Conversely, for 
$\theta=80.0$ used in the model, the corresponding $\beta_{\rm j}=0.92^{+0.08}_{-0.11}$.

\subsection{3C\,309.1}

We use the position measurements of Aaron (1996 and priv. comm.), and 
adopt the jet geometry as modelled by Kus (1993). Only upper limits are available for
the core offset at frequencies higher than 5\,GHz, as the separation between the 
core and closest optically thin feature remains constant. 
The derived $\Omega_{r\nu}$ decrease substantially towards higher frequencies. We 
take it as evidence for larger $k_{\rm r}$ due to increasing opacity at shorter distances
from the jet origin. Between 1.5 and 2.3\,GHz, $k_{\rm r}$ is likely to be close to unity,
so we take the corresponding derived quantities as best estimates of 
physical conditions in the jet. The estimated $N_1 = 17000$ is larger than
in the other objects listed in Table \ref{tb:jet-model}, possibly implying a 
stronger pressure confinement in 3C\,309.1. The latter can be reconciled with the 
results from optical spectroscopy
(Forbes et al. 1990) suggesting that 3C\,309.1 may have a massive 
($\dot{M} \ge 1000$\,M{\solar}yr$^{-1}$) cooling flow within a radius of 
11.5$h^{-1}$\,kpc.

\begin{figure}[t]
\cl{\psfig{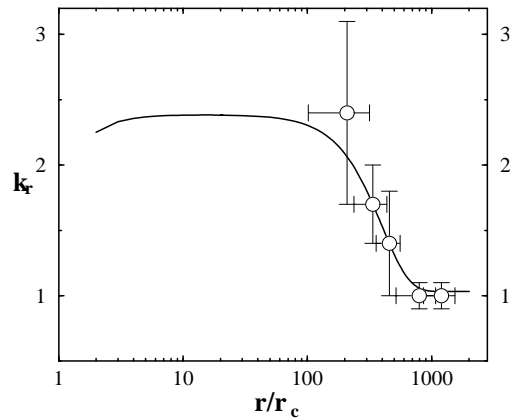}}
\caption{Opacity in the jet of 3C\,309.1. Circles are the measured
values of $k_{\rm r}$ in 3C\,309.1 at different frequencies; solid
line shows changes of $k_{\rm r}$ due to pressure gradients described
in section 2.3. The model parameters are the same as in
Fig. 2.\label{fg:309kr}}
\end{figure}

The variations of $k_{\rm r}$ necessary to account for the difference in measured
$\Omega_{r\nu}$ are compared in Fig. \ref{fg:309kr} with changes due to the pressure
gradients described in section \ref{sc:pressure}.
Here we assign $r_{\rm core}(1.5\,{\rm GHz}) = 1200r_{\rm c}$, treat upper limits of 
$\Omega_{r\nu}$ as measured points, and use the same model parameters
as in Fig. \ref{fg:blrmod}. 
The behavior of jet opacity  displayed in Fig. \ref{fg:309kr} is in agreement 
with the measurements of Aaron et al. (1997), who report a moderately high
rotation measure $RM_{\rm core} \sim -1600$ with strong gradients across
a nuclear region of $\sim 3h^{-1}$\,pc in size.
It should be stressed however that the model shown in Fig \ref{fg:309kr} 
is used for purely
illustrative purposes, and we do not attempt to make quantitative conclusions or
rule out other explanations for the observed core offsets in 3C\,309.1. For instance,
external free--free absorption in the BLR clouds may result in similar changes of 
$k_{\rm r}$. 

\subsection{3C\,345 \label{sc:345jet}}

Properties of the ultracompact jet in 3C\,345 have been discussed recently 
in several papers, based on X--ray (Unwin et al. 1994, 1997),
VLBI (Zensus et al. 1995), and multifrequency observations of the source
(Bregman et al. 1986; Webb et al. 1994; Stevens et al. 1996). We adopt a
spectral index $\alpha_{\rm j} = -0.6$ (Unwin et al. 1994), and derive the geometric
parameters for our model from the statistical properties of superluminal 
features embedded in the extended jet (Zensus et al. 1995; Lobanov 1996).

To obtain a plausible estimate of $\delta_{\rm j}$, one can assume that the jet 
carries the least kinetic power, so that $\gamma_{\rm j} = (1+\beta^2_{\rm app})^{0.5}$ 
(the assumption is 
valid only if the observed speeds are those of bulk motions in the jet). 
Under this assumption, the variations of
viewing angle are derived for the jet components in the immediate vicinity of
the core (Lobanov 1996); and we estimate $\gamma_{\rm j} \approx 5.6$ and
$\theta_{\rm j}\approx7.6\deg$ for the Lorentz factor and viewing 
angle of the ultracompact jet in 3C\,345. These values are close to the fitted
values from the synchrotron self--Compton (SSC) models applied to 3C\,345 (Unwin
et al. 1994; Webb et al. 1994). 

The jet opening angle, $\phi_{\rm j} \approx 2.4\deg$, is calculated from measuring the
jet size within 1\,mas distance from the core. The particle density 
$N_1=1300$\,cm$^{-3}$ is almost 20 times smaller than the value given in Unwin et al.
(1994) for an SSC model of the VLBI core. Zensus et al. (1995) have indicated that the
SSC value of $N_1$ might be roughly 10 times smaller, to accommodate the
magnetic field and particle density found in the jet superluminal features. The latter
prediction is consistent with our estimate of $N_1$  within a factor of 2.

\begin{table}[t]
\caption{Derived parameters of the ultracompact jet \label{tb:mfield}}
\footnotesize
\begin{center}
\medskip
\begin{tabular}{cccc} \hline\hline
$\nu$ & $B_{\rm core}$ & $\Delta r_{\rm pc}$ & $\Delta r_{\rm mas}$ \\
${\rm [GHz]}$ & ${\rm [G]}$ & & \\ \hline
5.0 & $0.09\pm 0.05$ & $16.3\pm 4.6$ & $0.57\pm 0.16$ \\
8.4 & $0.15\pm 0.08$ &  $9.7\pm 2.8$ & $0.34\pm 0.10$ \\
10.6 & $0.18\pm 0.10$ & $7.7\pm 2.2$ & $0.27\pm 0.08$ \\
22.2 & $0.38\pm 0.21$ & $3.7\pm 1.0$ & $0.13\pm 0.04$ \\\hline
\end{tabular}
\end{center}
Notes: the angular offsets  are given for a jet with $\theta=7.6^{\deg}$.
\end{table}

The calculated offset measures in 3C\,345 given  in Table \ref{tb:jet-model} 
decrease at lower frequencies. The corresponding $k_{\rm r}$ must decrease by about
20\%, between 5 and 11\,GHz (which is within the errors of the fit given in
Fig. \ref{fg:offset}). Assuming that $k_{\rm r}(5\,{\rm GHz}) = 1$, the decrease
in $\Omega_{r\nu}$ corresponds to $k_{\rm r}(11\,{\rm GHz}) = 0.81\pm0.14$. This implies
that the optical depth in the jet also becomes smaller at higher frequencies ---
a situation that does not have a simple explanation in the frameworks of both
synchrotron self--absorption and external free--free absorption. With this 
argument in mind, and also considering the magnitude of errors in
the derived offset measures, it appears more likely that the decrease of 
$\Omega_{r\nu}$ results from blending effects due to the limited resolution. With decreased
angular resolution at lower frequencies, the blending between the core and a nearest
jet feature (remaining unidentified in the VLBI maps) should 
become progressively stronger, resulting in a systematic trend in $\Omega_{r\nu}$.
However, we cannot
entirely exclude an explanation involving peculiar physical conditions in the source.
We therefore take the average $\Omega_{r\nu} = 10.7\pm5.4$ as the reference
value for 3C\,345.

The corresponding average characteristic magnetic field, $B_1 = 1.4\pm0.7$\,G, is 
lower than most of the values obtained by Webb et al. (1994), but can be reconciled 
with the results from Unwin et al. (1994), considering the difference of the
adopted particle densities.

The average total power of the jet is
$L_t \approx (1.5\pm0.4)
\cdot 10^{47}$\,erg\,s$^{-1}$, and we find $L_{\rm syn} 
\approx (1.8\pm0.5)\cdot 10^{46}$\,erg\,s$^{-1}$. 
With the derived magnetic field, the maximum brightness temperature is
$T_{\rm b\,max} \approx 2.5 \cdot10^{12}$\,K. The obtained value
is consistent with the $10^{12}$\,K inverse Compton limit, similarly
to the mean values of the maximum brightness temperature in a
sample of superluminal sources \cite{vc94}. Arguments based on the
hypothesis of equipartition between relativistic particle and magnetic
energy density ($\varepsilon_{\rm p}/\varepsilon_{\rm B} \sim 1$) predict
a lower value  of maximum brightness temperature $T_{\rm b\,eq} \approx 4.5\cdot
10^{11}$\,K, Readhead 1994). The relativistic particle energy density,
$\varepsilon_{\rm p} \propto T_{\rm b}^{4.5}$, and the energy density in the
magnetic field, $\varepsilon_{\rm B} =B^2/8\pi \propto T_{\rm b}^{-4}$.  This gives
$\varepsilon_{\rm p}/\varepsilon_{\rm B} \approx (T_{\rm b\,max}/T_{\rm b\,eq})^{8.5}
\approx 2\cdot 10^6$ for the
ultracompact jet, and therefore the jet must lose its energy through
X-ray emission due to inverse Compton scattering. Unwin et al. (1994)
argue that physical conditions in the extended jet may as well deviate 
significantly from the equipartition.
According to Unwin et al. (1994), $\varepsilon_{\rm p}/\varepsilon_{\rm B} \sim 10^2-10^4$
at $r\sim10$--$100$\,pc, which can be a natural consequence of evolution and
radiation losses in a plasma with original $\varepsilon_{\rm p}/\varepsilon_{\rm B} \sim 10^6$.

\begin{figure}[t]
\cl{\psfig{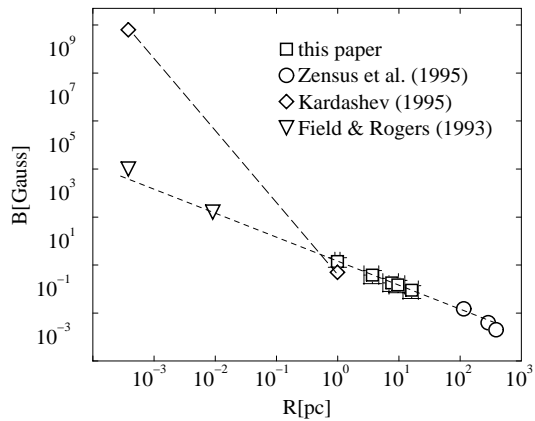}}
\caption{Magnetic field distribution in 3C\,345. Squares show the
magnetic field in the compact jet derived from the
frequency dependent shift 
of the core. Circles are the homogeneous synchrotron model estimates 
of magnetic field in the extended jet components (Zensus et al. 1995). 
Triangles show the characteristic magnetic field values from a model of
magnetized accretion disk (Field and Rogers,
1993). Diamonds are the theoretical estimates from Kardashev's (1995)
model for the dipole magnetic field around a supermassive rotating black
hole. \label{fig:mfield}}
\end{figure}

\subsection{Magnetic field in the jet of 3C\,345}

Table \ref{tb:mfield} summarizes the derived positions and magnetic
fields in the VLBI core of 3C\,345 at different frequencies. 
According
to the derived positions, the location of the core observed at 22\,GHz
should be about 4\,pc away from the jet origin. 
The angular offsets are all smaller than
the VLBI beam size at corresponding frequency; therefore, the central engine
is likely to be blended with the VLBI core even if the opacity is low
enough for the emission from the innermost regions of the jet to be
observed. Under such circumstances, evidence for emission from the
inner jet can only be found in the spectral index or turnover
frequency maps (Lobanov et al. 1997).

In Fig. \ref{fig:mfield}, the derived magnetic field is compared
with the values obtained from synchrotron emission calculations for
the extended jet (Zensus et al. 1995), as well as with 
model predictions for the central engine in 3C\,345.
The theoretical values of magnetic field in the close vicinity of
central engine are calculated from a model of 
thin, magnetically driven accretion disk (Field and Rogers 1993), and from a
model of supermassive black hole surrounded by a strong magnetic
field \cite{kar95}.
The accretion disk model predicts $B\approx 10^4$ at $r\approx R_{\rm g}$, and
$B\approx 160$\,G at $r\approx 24R_{\rm g}$ ($R_{\rm g}=2GM_{\rm bh}/c^2$ is the 
Schwarzschild radius of the central
black hole). The mass of central black hole can be related to $B_1$, so that
$M_{\rm bh} \approx 2.7\cdot 10^9 B_1$M\solar; and in 3C345: 
$M_{\rm bh} \approx 4\cdot 10^9$M\solar.
Kardashev (1995) suggests that a strong
dipole magnetic field may exist in the vicinity of a supermassive black
hole. The maximum field strength can be estimated from the
equilibrium relations $M_{\rm bh}\sim M_{\rm d} \sim M_{\rm B}$, where $M_{\rm bh}$,
$M_{\rm d}$, $M_{\rm B}$ are masses of black hole, its disk, and its magnetic
field respectively. Then $B_{\rm max}=c^4 G^{-1.5}M^{-1}_{\rm bh} = 2.4\cdot 10^{19}
(M_{\rm bh}/{\rm M}\solar)^{-1}$. This gives
for 
3C\,345: $B_{\rm max}\approx 6.3\cdot 10^9$\,G. For a dipole magnetic field, one obtains
$B\propto r^{-3}$), so that $B\approx0.8$\,G at
1\,pc, slightly lower than our estimate in Table \ref{tb:jet-model}.  
To match the derived $B_1$, the mass of central black hole in Kardashev's model
must be $M_{\rm bh} = (c B^{1/2} r^{3/2})/(2\sqrt{2} G^{3/4}) \approx 7\cdot 10^9 B^{1/2}
r^{3/2} {\rm M}\solar \approx 8.3\cdot10^9\,{\rm M}\solar$.

\subsection{3C\,395 and 4C\,39.25}

The core offsets are known from geodetic VLBI measurements between 2.3 and 8.4\,GHz
(Lara et al. 1996; Guirado et al. 1995). We use the models derived for the parsec--scale 
jets in these sources (Lara et al. 1994; Alberdi et al. 1993), and adopt the equipartition
values of $N_1$. The resulting $\Omega_{r\nu}$ are
similar to the 2.3--5\,GHz offset measure in 3C\,309.1.  Other derived parameters are
also comparable to their counterparts in 3C\,345 and 3C\,309.1. The derived 
synchrotron luminosity of 4C\,39.25 is quite high, compared to
$L_{\rm syn}$(10--90\,GHz)$=5.5\cdot 10^{44}$\,erg\,s$^{-1}$ given in Bloom \& Marscher (1991), 
even taking into account the spectral limits.

Recent observations of 4C\,39.25 at 22.2 and 43.2\,GHz (Alberdi et al. 1997) indicate
no detectable position shift in the source. From the measured $\Omega_{r\nu}$ between
2.3 and 8.4\,GHz, the expected shift between 22.2 and 43.2\,GHz is less than 0.05\,mas.
The latter value is below the image resolution at both frequencies ($\approx 0.3$ and
$\approx 0.17$\,mas). 

\subsection{1038+528\,A}

Marcaide \& Shapiro (1984) reported a 0.7\,mas shift between 2.3 and 8.4\,GHz in 
1038+528\,A. Subsequently, Marcaide et al. (1985) derived $0.5<k_{\rm r}<1.4$, based on 
VLBI data at 1.4, 2.3, 8.4, and 10.6\,GHz. Marcaide,
El\'osegui \& Shapiro (1994) applied a relativistic shock model by Gomez et al. (1993)
to explain the observed shift. Rioja \& Porcas (1997) have made position measurements
at 2.3, 8.4, and 15\,GHz, and argued that at least a fraction of the observed offset
can result from limited resolution and blending at lower frequencies. The 
$\Omega_{r\nu}$(2.3--8.4\,GHz) calculated from the data of Rioja et al. (1997) is
similar to the values obtained for 4C\,39.25 and 3C\,395 (although the projection 
effects should be kept in mind, in view of the different viewing angles in these
sources). The upper limit given for the offset between 8.4 and 15\,GHz corresponds to
$k_{\rm r} \ge 4.4$, making it rather difficult to be explained by synchrotron 
self--absorption.
Even for free--free absorption, it would require very strong density gradients 
($n\ge 4$) in the absorbing medium, in order to reproduce the given limit on $k_{\rm r}$.

\begin{figure}[t]
\cl{\psfig{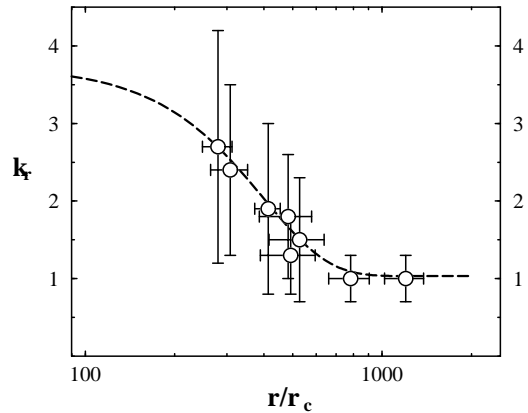}}
\caption{Relative changes of opacity in the ultracompact
jets. Reference point is the 1.5--2.3\,GHz offset in 3C\,309.1, with
$k_{\rm r} =1$ and $r = 1200r_{\rm c}$ assigned to it. For the rest of
the data, the $k_{\rm r}$ and $r$ are calculated from the
corresponding $\Omega_{r\nu}$ measured in the respective source rest
frames.  Dashed line is a model similar to that described in section
2.3, but with $a(r_{\rm c})=6$.\label{fg:alljets}}
\end{figure}

\section{Opacity in the jets}

Assuming that the frequency term in (\ref{eq:theor3}) dominates core 
position offsets, the measured
$\Omega_{r\nu}$ can be used for comparing the relative opacity in different sources.
To make such a comparison, the offset measures and observing frequencies must be
transformed
to the respective source rest frames: $\Omega^{\prime}_{r\nu} = \Omega_{r\nu} \sin\theta_{\rm j}
(1+z)$; $\nu^{\prime} = \nu (1+z)$. We then postulate that
the largest $\Omega^{\prime}_{r\nu}$ corresponds to $k_{\rm r}=1$, $r=1200r_{\rm c}$, and use it as 
a reference point for calculating the $k_{\rm r}$ for the rest of the offset measurements. 
The results are plotted in Fig. \ref{fg:alljets}. In 3C\,395, the calculated 
$k_{\rm r}$ is very large ($k_{\rm r}>17$), suggesting that the jet viewing angle may be greater 
than the value cited by Lara et al. (1994).
The apparent increase of $k_{\rm r}$ at shorter radial distances is consistent with the
self--absorption scenario described in section \ref{sc:pressure} 
(the measured $\Omega^{\prime}_{r\nu}$ become smaller at higher
frequencies at which the opacity is larger).
The linear scale in Fig. \ref{fg:alljets} serves only as an illustration,
since the intrinsic properties of the sources (reflected by the $B_1^{k_{\rm b}}F$ term) 
may be different. 
Detailed studies of each individual object
are required for establishing a connection between the jet linear scale and
opacity properties.
A general prediction is that the position offset should be larger in flat--spectrum
cores in which $k_{\rm r}\sim 1$. When $k_{\rm r}$ increases, the offsets are expected to become
smaller, and the core spectrum should be inverted.

\begin{figure}[t]
\cl{\psfig{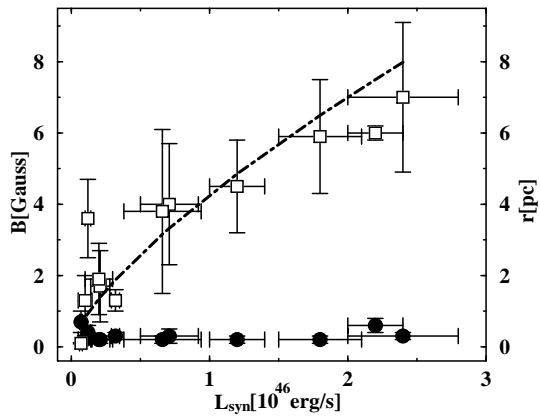}}
\caption{Magnetic field (filled circles) and distance of the core
(squares) with respect to the jet synchrotron luminosities. The data
are shown for $\nu^{\prime} = 22$\,GHz.  Dot--dashed line is $r_{\rm
core} \propto L_{\rm syn}^{2/3}$.
\label{fg:corejets}}
\end{figure}

A reasonably good agreement found in Fig. \ref{fg:alljets} between the derived and
model $k_{\rm r}$ indicates that synchrotron self--absorption may be responsible
for the observed properties of optically thick emission from ultracompact jets. One can
expect then to find similar physical conditions in the VLBI cores of different sources,
if the measurements are done at the same rest frame frequency. To illustrate this, 
we calculate the magnetic fields and core distances at $\nu^{\prime} = 22$\,GHz, and
plot them versus jet synchrotron luminosity (Fig. \ref{fg:corejets}). The 
derived $r_{\rm core}$ increase, with increasing $L_{\rm syn}$, and follow roughly the proportionality
$r_{\rm core}\propto L_{\rm syn}^{2/3}$ resulting from (\ref{eq:theor5}) and (\ref{eq:theor7}). 
The corresponding magnetic field however remains nearly constant, with an average
of $B_{\rm core} = 0.30\pm0.08$\,[G].
One may expect the values of magnetic field to vary stronger, and exhibit a rather large
scatter, if jet emission is affected by external factors such as free--free absorption.
The homogeneity of $B_{\rm core}$ seen in Fig. \ref{fg:corejets} suggests that the jet
plasma is a primary factor determining the location and properties of VLBI cores,
and that intrinsic physical conditions in the jets must be fairly similar.
If this suggestion holds for other sources, the expected
magnetic field in VLBI core should be of the order of
$B_{\rm core}(\nu_{\rm obs}) \approx 0.3\left[ 0.045  \nu_{\rm obs} (1+z) \right]^{m/k_{\rm r}}$\,[G],
for $\nu_{\rm obs}$ measured in GHz.

\section{Conclusions}

The frequency dependent position shift of VLBI cores can be used for studying the 
most compact regions of extragalactic jets, and obtaining quantitative estimates 
for physical conditions in the immediate vicinity of central engine. 
In the absence of high--precision position measurements, a more simplistic approach can
be employed, assuming that positions of optically thin features in the jet remain 
the same at all frequencies. Then, the position offsets of optically thin features
can be interpreted as a frequency dependent shift of the self--absorbed core of the jet.
In some sources (1038+528\,A, 3C\,345, Cygnus A), insufficient resolution and 
blending at lower frequencies may undermine the offset measurements. A way to detect
and, to a certain degree, to correct the errors due to the blending can be found in 
introduction
of an offset measure, $\Omega_{r\nu}$, which is supposed to remain constant for the
case of ideal measurements and unchanged opacity of the jet. Listed below are the main 
results from deriving the offset measures in 6 sources with reported position offsets
of VLBI core:

1.~The jet luminosity, the maximum brightness temperature, the core magnetic field 
and location with respect to the jet origin can be determined, from offset measures. 
For the well studied sources (Cygnus A,
3C\,345), the derived values are shown to agree well with values determined by
other methods. In Cygnus A, the jet--to--counter--jet ratio determined from the 
estimated location of the central engine results in a self--consistent source geometry.
In 3C\,345, the derived maximum brightness temperature indicates that the ultracompact jet
is strongly particle--dominated, and must release its energy through inverse Compton 
scattering. The obtained magnetic field distribution in the ultracompact jet of 3C\,345 
is in a good
agreement with the values derived from inhomogeneous synchrotron models applied to the
extended jet components. Based on the derived characteristic magnetic field, it appears
more likely that the jet is produced by a thin, magnetized accretion disk, rather than
by a single supermassive black hole with a strong dipole magnetic field.

2.~External pressure and density gradients typical for the broad line
region may change the optical depth along the jet via both
synchrotron self--absorption and free--free absorption in an ambient
medium.  The offset measures derived for different frequency pairs and
different objects can be used for studying the opacity effects in
ultracompact jets. Changes of jet opacity in 3C\,309.1 deduced from the
variations of $k_{\rm r}$ appear to be consistent with a self--absorbed jet
propagating through a region with strong density gradients. Similar
conclusion can be drawn from the variations of $k_{\rm r}$ determined for
all studied objects, although the agreement with the model may in this case 
be coincidental (one argument for making such a reservation is that 
the initial pressure gradients required to fit the data for all sources
simultaneously are very high: $N(r) \propto r^{-6}$, at $r\sim r_{\rm c}$).

3.~The distance at which the observed VLBI core is located at a given frequency is
scaled with jet luminosity, so that $r_{\rm core}\propto L_{\rm syn}^{2/3}$. At the
same time the magnetic field in the core appears to be almost unchanged,
from one source to another. This can be taken as another evidence for 
the synchrotron self--absorption to be a primary factor influencing
the observed properties of the jet core.

Further, more detailed studies of absorption in ultracompact jets are required 
for confirming the conclusions stated above, and for constructing better models
of individual sources. Nearly simultaneous, multifrequency VLBI data obtained
with phase--referencing should be most useful for a reliable determination of the
core offsets, and subsequent studies of the most compact regions in active galactic
nuclei.

\section*{Acknowledgements}
We would like to thank S.~Aaron, T.~Krichbaum, R.~Porcas and M.~Rioja for fruitful
discussions and providing the unpublished data for this study.


\begin{thebibliography}{}

 \bibitem [(Aaron 1996)]{aar96} Aaron, S.E. 1996, Ph.D. Thesis, Brandeis University

\bibitem[(Aaron et al. 1997)] {aar+97} Aaron, S.E., Wardle, J.F.C, and 
Roberts, D.H. 1997, Vistas in Astronomy, 41, 225

\bibitem[(Alberdi et al. 1993)]{alb+93} Alberdi, A., Marcaide, J.M., Marscher, A.P.
et al. 1993, ApJ, 402, 160

\bibitem[(Alberdi et al. 1997)]{alb+96} Alberdi, A., Krichbaum, T.P., Graham, D.A.,
et al. 1997, A\&A (in press)

\bibitem[(Baldwin et al. 1996)] {bal+96} Baldwin, J.A., Ferland, G.J., Korista, K.T.
et al. 1996, ApJ, 461, 664

\bibitem[(Beasley \& Conway 1995)]{bc95} Beasley, A.J. \& Conway J.E. 1995,
in Very Long Baseline Interferometry and the VLBA, eds. J.A. Zensus, 
P.J. Diamond, \& P.J. Napier (Cambridge, Cambridge University Press), 291

\bibitem[(Begelman 1995)]{beg95} Begelman, M.C. 1995, in Publications 
of the National Academy of Sciences, Vol.92, Quasars and Active Galactic 
Nuclei: High Resolution Radio Imaging, eds. M.H. Cohen \& K.I. Kellermann
(reprint: Charlottesville, NRAO), 11442

\bibitem[(Begelman et al. 1984)]{bbr84} Begelman, M.C., 
Blandford, R.D., \& Rees, M.J. 1984, Rev. Mod. Phys., 56, 255

\bibitem[(Biretta et al. 1986)]{bmc86} Biretta, J.A., Moore, R.L., \& Cohen,
M.H. 1986, ApJ, 308, 93

\bibitem[(Blandford \& K\"onigl 1979)]{bk79} Blandford, R.D. \& K\"onigl, A. 1979, ApJ, 232, 34.

\bibitem[(Blandford \& Rees 1974)]{br74} Blandford, R.D. \& Rees, M.J. 1974, MNRAS,
169, 395

\bibitem[(Bloom \& Marscher 1991)] {bm91} Bloom, S.D. \& Marscher, A.P. 1991, ApJ,
366, 16

\bibitem[(Blumenthal \& Gould 1970)]{bg70} Blumenthal, G.R. and Gould, R.J.
1970, Rev. Modern Phys., 4, 237

\bibitem[(Bregman et al. 1986)]{bre+86} Bregman, J.N., Glassgold, A.E., 
Huggins, P.J., et al. 1986, ApJ, 301, 708

\bibitem[(Carilli \& Barthel)]{cb96} Carilli, C.L. \& Barthel, P.D. 1996, A\&A Rev.,
7, 1

\bibitem[(Cassidy \& Raine 1993)]{cr93} Cassidy, I. \& Raine, D.J. 1993, MNRAS, 260,
385

\bibitem[(Daly \& Marscher 1988)]{dm88} Daly, R.A., Masrcher, A.P. 1988, ApJ, 334,
539

\bibitem[(Ferguson et al. 1997)]{fkf97} Ferguson, J.W., Korista, K.T.,
\& Ferland, G.J. 1997, ApJS, 110, 287

\bibitem[(Field \& Rogers 1993)] {fr93} Field, G.B. \& Rogers,  R.D. 1993, ApJ, 403, 94

\bibitem[(Fomalont 1995)]{fom95} Fomalont, E. 1995, in Very Long Baseline 
Interferometry and the VLBA, eds. J.A. Zensus, P.J. Diamond, \& P.J. Napier
(Cambridge, Cambridge University Press), 205

\bibitem[(Georganopoulos \& Marscher 1996)] {gm96} Georganopoulos, M. \& 
Marscher, A.P. 1996 in ``Energy Transport in Radio Galaxies and Quasars'',
eds. P.E.~Hardee, A.H.~Bridle, J.A.~Zensus, ASP Conference Series, v.100, 67

\bibitem[(Gomez, Alberdi, \& Marcaide 1993)]{gam93} Gomez, J.L., Alberdi, A., 
\& Marcaide, J.M. 1993 A\&A 274, 55

\bibitem[(Ghisellini et al. 1992)]{ghi+92} Ghisellini, G., Celotti, A., 
George, I.M., \&Fabian, A.C. 1992, MNRAS, 258, 776

\bibitem[(Guirado et al. 1995)]{gui+95} Guirado, J.C., Marcaide, J.M.,
Alberdi, A., et al. 1995, AJ, 110, 2586

\bibitem[(Jones et al. 1996)]{jon96} Jones, D.L., Tingay, S.J., Murphy, D.W., 
et al. 1996, ApJ, 466, 63

\bibitem[(Hummer 1988)]{ham88} Hummer, D.G. 1988, ApJ, 327, 477

\bibitem[(Kaspi et al. 1996)]{kas+96} Kaspi, S., Smith, P.S., Maoz, D., Netzer, H.,
Januzzi, B.T. 1996, ApJ, 471, L75

\bibitem[(Ko\"onigl 1981)]{kon81} K\"onigl, A. 1981, ApJ, 243, 700

\bibitem[(Krichbaum et al. 1997)]{kri+97} Krichbaum, T.P., Alef, W., Witzel, A., 
Zensus, J.A., Booth, R.S., Greve, A., \& Rogers, A.E.E. 1997 A\&A, in press

\bibitem[(Krichbaum \& Witzel 1992)]{kw92} Krichbaum, T.P. \& Witzel, A. 1992, in
``Variability of Blazars'', eds. E.~Valtaoja \& M.~Valtonen (Cambridge: CUP), p.205

\bibitem[(Kardashev 1995)]{kar95} Kardashev, N.S. 1995, MNRAS, 276, 515

\bibitem[(Kus 1993)] {kus93} Kus, A.J. 1993, in ``Sub--arcsecond Radio Astronomy'',
eds. R.J.~Davis \& R.S.~Booth (Cambridge: CUP), p.365

\bibitem[(Lara et al. 1994)]{lar+94} Lara, L., Alberdi, A., Marcaide, J.M.,
\& Muxlow, T.W.B. 1994, A\&A, 285, 393

\bibitem[(Lara et al. 1996)]{lar+96} Lara, L., Marcaide, J.M., Alberdi, A.,
\& Guirado, J.C. 1996, A\&A, 314, 672

\bibitem[(Leahy 1991)] {lea91} Leahy, J.P. in ``Beams and Jets in Astrophysics'',
ed. P.A.~Hughes (Cambridge: CUP), p.100

\bibitem[(Levinson et al. 1995)]{lev+95} Levinson, A., Laor, A., \& Vermeulen, R.C.
1985, ApJ, 448, 589

\bibitem[(Lobanov 1996)]{lob96} Lobanov, A.P. 1996, Ph.D. Thesis (New Mexico 
Institute of Mining and Technology, Socorro NM)

\bibitem[(Lobanov et al. 1997)]{lcz97} Lobanov, A.P., Carrara, E., \& Zensus, J.A.
1997, Vistas in Astronomy, 41, 253

\bibitem[(Lovelace \& Romanova 1996)] {lr96} Lovelace, R.V.E. \& Romanova, M.M. 1996,
in ``Cygnus A -- Study of a Radio Galaxy'', eds. C.L.~Carilli \& D.E.~Harris 
(Cambridge: CUP), p.98

\bibitem[(Marcaide \& Shapiro 1984)]{ms84} Marcaide, J.M. \& Shapiro, I.I.
1984, ApJ, 276, 56

\bibitem[(Marcaide et al. 1985)]{mar+85} Marcaide, J.M., Shapiro, I.I., Corey, B.E.,
        et al. 1985, A\&A, 142, 71

\bibitem[(Marcaide et al. 1994)]{mes94} Marcaide, J.M., El\'osegui, P., \&
Shapiro, I.I. 1994, AJ, 108, 368

\bibitem[(Marscher 1980)] {mar80} Marscher, A.P. 1980, ApJ, 235, 386

\bibitem[(Marscher 1995)]{mar95} Marscher, A.P. 1995, in Publications 
of the National Academy of Sciences, Vol.92, Quasars and Active Galactic 
Nuclei: High Resolution Radio Imaging, eds. M.H. Cohen \& K.I. Kellermann
(reprint: Charlottesville, NRAO), 11442

\bibitem[(Readhead 1994)]{rea94} Readhead, A.C.S. 1994, ApJ, 426, 51

\bibitem[(Rioja \& Porcas 1997)] {rp+97} Rioja, M.J. \& Porcas, R.W. 1997,
in Proceedings of IAU Cooloquium No.\,164, eds. J.A.~Zensus, J.M.~Wrobel \& G.B.~Taylor
(Cambridge: CUP)

\bibitem[(Rybicki \& Lightman 1979)] {rl79} Rybicki, G.B. \& Lightman, A.P. 1979, 
``Radiative Processes in Astrophysics'', (J.Wiley: New York)

\bibitem[(Scheuer 1960)] {sch60} Scheuer, P.A.G. 1960, MNRAS, 120, 231


\bibitem[(Sol et al 1989)] {spa89} Sol, H., Pelletier, H. \& Ass\'{e}o, E. 1989, MNRAS,
237, 411

\bibitem[(Stevens et al. 1996)] {ste+96} Stevens, J.A., Litchfield, S.J., 
Robson, E.I., et al. 1996, ApJ, 466, 158

\bibitem[(Udomprasert et al. 1997)]{udo+97} Udomprasert, P.S., Taylor, G.B.,
Pearson, T.J., \& Roberts, D.H. 1997, ApJ, 483, L1

\bibitem[(Unwin et al. 1994)]{unw+94} Unwin, S.C., Wehrle, A.E., Urry, C.M., 
et al. 1994, ApJ, 432, 103

\bibitem[(Unwin, et al. 1997)] {unw+97} Unwin, S.C., Wehrle, A.E., Lobanov, A.P.,
et al. 1997, ApJ, 480, 596

\bibitem[(Vermeulen \& Cohen 1994)]{vc94} Vermeulen, R.C. \& Cohen, M.H.
1994, ApJ, 430, 467

\bibitem[(Vermeulen, Readhead \& Backer 1994)]{vrb94} Vermeulen, R.C., Readhead, A.C.S.,
\& Backer, D.C. 1994, ApJ, L41.

\bibitem[(Walker, Romney \& Benson)]{wrb94} Walker, R.C., Romney, J.D., \& 
Benson, J.M. 1994, ApJ, 429, L45

\bibitem[(Wardle et al. 1994)]{war+94} Wardle, J.F.C., Cawthorne, T.V., Roberts, D.H., \& Brown, L.F. 1994, ApJ, 437, 122

\bibitem[(Webb et al. 1994)] {web+94} Webb, J.R., Shrader, C.R., Balonek, T.J., et
al. 1994, ApJ, 422, 570

\bibitem[(Zensus, Cohen \& Unwin 1995)]{zcu95}Zensus, J.A., Cohen, M.H., 
\& Unwin, S.C. 1995, ApJ, 443, 35

\bibitem[(Zensus et al. 1997)]{zen+97} Zensus, J.A. et al. 1997, (in preparation)

\end{thebibliography}
\end{document}